\RequirePackage{amsmath} 
\RequirePackage{array}

\documentclass[12pt]{iopart}

\usepackage{lmodern}					
\usepackage[T1]{fontenc}
 
\usepackage[cm]{sfmath}
\usepackage[utf8]{inputenc}
\usepackage[a4paper,hmargin=2.5cm,vmargin=2.25cm]{geometry}

\usepackage{float}
\usepackage{iopams}
\usepackage{pgfplots}
\usepackage{lscape}
\usepackage{adjustbox}
\PassOptionsToPackage{hyphens}{url}\usepackage{hyperref}
\hypersetup{
	colorlinks=false,
	linkcolor=black,
	filecolor=blue,      
	urlcolor=black,
}

\usepackage[english]{babel}

\usepackage{amssymb}
\usepackage{amsfonts}
\usepackage{mathtools}
\usepackage[binary-units=true]{siunitx}
\usepackage{caption}
\usepackage{subcaption}
\usepackage{graphicx}
\usepackage{xcolor, colortbl}

\usepackage{booktabs}
\usepackage{tabularx}

\usepackage{tikz}
\usetikzlibrary{external,automata,trees,shadows,matrix,fit,decorations,positioning,shapes,plotmarks,patterns,arrows,calc,quotes,tikzmark,arrows.meta}

\renewcommand{\mat}[1]{\ensuremath{{\mathbf{\MakeUppercase{#1}}}}}

\makeatletter
\renewcommand{\vec}[1]{%
	\ifcat\relax\noexpand#1%
	\ensuremath{\boldsymbol{\lowercase{#1}}}%
	\else
	\ensuremath{\mathbf{\lowercase{#1}}}%
	\fi
}
\makeatother

\makeatother

\makeatletter

\newcommand{\transpose}[1]{\ensuremath{{#1}^{\textsc{t}}}}

\newcommand{\inverse}[1]{\ensuremath{{#1}^{-1}}}

\newcommand{\R}{\ensuremath{\mathbb{R}}}

\newcommand{\norm}[1]{\left|\left|#1\right|\right|}

\usepackage{pifont}

\usepackage[font=small,labelfont=bf]{caption} 
\usepackage[noadjust]{cite}

\begin{document}
	
	\title[]{Linear stimulus reconstruction \emph{works} on the KU Leuven audiovisual, gaze-controlled auditory attention decoding dataset}
	\author{Simon Geirnaert$^{1,2,3}$, Iustina Rotaru$^{1,2}$, Tom Francart$^{2,3}$, Alexander Bertrand$^{1,3}$}
	\address{\normalfont$^1$KU Leuven, Department of Electrical Engineering (ESAT), Stadius Center for Dynamical Systems, Signal Processing and Data Analytics}
	\address{\normalfont$^2$KU Leuven, Department of Neurosciences, ExpORL}
	\address{\normalfont$^3$Leuven.AI - KU Leuven institute for AI,}
	\ead{simon.geirnaert@kuleuven.be, alexander.bertrand@kuleuven.be\footnotetext{\normalfont Financial support was provided by the Research Foundation Flanders (FWO) (junior postdoctoral fellowship fundamental reserach 1242524N for S. Geirnaert, SBO mandate 1S14922N for I. Rotaru, and FWO project G0A4918N) and by the European Research Council (ERC) under the European Union’s research and innovation programme (grant agreement No 802895 and grant agreement No 101138304). Views and opinions expressed are however those of the author(s) only and do not necessarily reflect those of the European Union or ERC. Neither the European Union nor the granting authority can be held responsible for them.}}
	
	\begin{abstract}
		\normalfont In a recent paper, we presented the KU Leuven audiovisual, gaze-controlled auditory attention decoding (AV-GC-AAD) dataset, in which we recorded electroencephalography (EEG) signals of participants attending to one out of two competing speakers under various audiovisual conditions. The main goal of this dataset was to disentangle the direction of gaze from the direction of auditory attention, in order to reveal gaze-related shortcuts in existing spatial AAD algorithms that aim to decode the (direction of) auditory attention directly from the EEG. Various methods based on spatial AAD do not achieve significant above-chance performances on our AV-GC-AAD dataset, indicating that previously reported results were mainly driven by eye gaze confounds in existing datasets. Still, these adverse outcomes are often discarded for reasons that are attributed to the limitations of the AV-GC-AAD dataset, such as the limited amount of data to train a working model, too much data heterogeneity due to different audiovisual conditions, or participants allegedly being unable to focus their auditory attention under the complex instructions. In this paper, we present the results of the linear stimulus reconstruction AAD algorithm and show that high AAD accuracy can be obtained within each individual condition and that the model generalizes across conditions, across new subjects, and even across datasets. Therefore, we eliminate any doubts that the inadequacy of the AV-GC-AAD dataset is the primary reason for the (spatial) AAD algorithms failing to achieve above-chance performance when compared to other datasets. Furthermore, this report provides a simple baseline evaluation procedure (including source code) that can serve as the minimal benchmark for all future AAD algorithms evaluated on this dataset.
	\end{abstract}
	
	\newpage
	\tableofcontents
	\markboth{}{}
	
	\section{Introduction}
	\label{sec:intro}
	Selective auditory attention decoding (AAD) methods aim to identify the sound source a person is attending to amidst a cocktail of sound sources based on neural recordings such as electro- or magneto-encephalography (EEG/MEG)~\cite{mesgarani2012selective,osullivan2014attentional}. A typical application of AAD is in cognitively-controlled hearing aids, which allow the user to steer their hearing aid towards the conversation they actually want to listen to, for example, in a cocktail party scenario~\cite{geirnaert2021eegBased}. 
	
	AAD methods can generally be classified into two different classes: stimulus decoding and direct classification~\cite{geirnaert2021eegBased,geirnaert2022phd}. \emph{Stimulus decoding} leverages so-called neural tracking: the brain tracks (features of) the attended stimulus better than other unattended stimuli~\cite{mesgarani2012selective,ding2012neural,golumbic2013mechanisms}. This class of algorithms, therefore, aims to reconstruct such features (e.g., the speech envelope) of the attended stimulus from the neural responses (backward modeling) to compare them with the presented stimuli to identify the correct speaker~\cite{osullivan2014attentional}. Alternatively, the neural responses can be predicted from each stimulus and compared with the actual neural responses recorded through EEG/MEG (forward modeling)~\cite{wong2018comparison}. Combinations of backward and forward modeling have also been proposed~\cite{decheveigne2018decoding}. The models used to reconstruct can vary from simple linear models to more complex non-linear models. This class of algorithms has been well-established in the literature, yielding robust results that have been countless times reproduced in various datasets~\cite{geirnaert2021eegBased}. The main disadvantage of this class is that the performance quickly degrades when less data is available to make a decision (i.e., at increasing decision speeds)~\cite{geirnaert2020interpretable}.
	
	The second class of \emph{direct classification} algorithms is becoming increasingly popular along with the rise of deep learning methods in EEG/MEG. While various approaches exist, a particularly popular subbranch consists of determining the \emph{spatial location} of the attended sound source solely based on the neural responses (spatial auditory attention decoding (Sp-AAD))~\cite{vandecappelle2021eeg,geirnaert2020fast,geirnaert2021riemannian,Su2022,pahuja2023xanet}. These neural responses are believed to reflect certain lateralization patterns based on the listening direction, offering an alternative approach to identifying the attended sound source based on their location relative to the listener. While the models used for Sp-AAD can be simple linear data-driven models (e.g., based on common spatial patterns~\cite{geirnaert2020fast}), there has been a recent surge in various non-linear deep learning models~\cite{vandecappelle2021eeg,Su2022,pahuja2023xanet}. Their main advantage is that they claim to require much less data to make an accurate decision about the attended speaker.
	
	However, in a recent paper~\cite{rotaru2024what}, we uncovered that these Sp-AAD algorithms suffer from various (unwanted) shortcuts or confounds in the datasets on which they are tested. One example of such a shortcut is trial fingerprints. We showed that Sp-AAD models are highly susceptible to these trial fingerprints and are (implicitly) overfitting due to high statistical similarity between segments in the training and test set. However, many papers resort to random cross-validation based on very short (and sometimes overlapping) segments as an evaluation scheme, leading to unrealistically and misleadingly high performance, as pointed out in Puffay et al.~\cite{puffay2023relating}. Generalizing across various trials or subjects turns out to be much harder than with stimulus decoding algorithms due to feature shifts across trials. A second shortcut arises from confounding signal components in the neural responses, such as eye movements. Various datasets have no or very limited restrictions on eye gaze (such as the KU Leuven AAD dataset of 2016~\cite{das2019dataset}), which could lead to congruent eye movements with the direction of auditory attention due to the tendency of listeners to (in)voluntarily direct their gaze toward the direction of the attended speaker. Data-driven models could then leverage eye movements to decode attention with a much higher accuracy, rather than the actual neural processes that are much harder to decode. While decoding eye movements for Sp-AAD might work just fine in many scenarios, it is not a fully robust way of decoding attention (e.g., when eavesdropping). Furthermore, it means that these models might not really decode attention directly from the brain, but instead exploit shortcuts in the EEG recordings that are in fact generated by the eyes, which (in many cases) turn out to be correlated with the direction of auditory attention.
	
	To uncover these potential shortcuts in Sp-AAD algorithms, Rotaru et al.~\cite{rotaru2024what} introduced a new two-speaker AAD dataset with forced congruent and incongruent spatial auditory and visual attention (i.e., the audiovisual, gaze-controlled auditory attention decoding dataset KU Leuven (AV-GC-AAD dataset)~\cite{rotaru_2024_11058711}). This AV-GC-AAD dataset is publicly available as a new benchmarking dataset. However, the original paper~\cite{rotaru2024what} did not include an analysis using the first class of stimulus decoding algorithms. In this paper, we want to provide such an analysis for various reasons. Firstly, as stimulus decoding algorithms are very robust and well-established, they allow to verify that the participants in the experiment followed the instructions and were indeed able to attend to the correct speaker, despite the varying visual attention cues that complicate focusing auditory attention. Secondly, it allows to provide a trustworthy baseline benchmark on the new dataset. As previously mentioned, many AAD datasets suffer from shortcuts and confounds regarding eye gaze. We have noticed that several Sp-AAD deep learning algorithms achieve skyrocketing performance on other datasets with minimal control for eye gaze while they fail on our new AV-GC-AAD dataset. However, AAD model developers often discard these failures and attribute them to an insufficient amount of training data, participants being unable to focus their auditory attention due to the distracting incongruent gaze-steering task, or a too-large variety across conditions. With this stimulus decoding analysis, we want to refute those arguments and, therefore, provide a minimal benchmark for future Sp-AAD and other algorithms.
	
	In Section~\ref{sec:dataset}, we give a brief overview of the AV-GC-AAD dataset, while the linear stimulus decoding algorithm for AAD is revisited in Section~\ref{sec:methods}. In Section~\ref{sec:experiments}, we explain the various experiments and evaluation setups and in Section~\ref{sec:results-discussion}, we show and discuss results. 
	
	\section{Dataset}
	\label{sec:dataset}
	The audiovisual, gaze-controlled auditory attention (AV-GC-AAD) dataset from KU Leuven was originally presented in Rotaru et al.~\cite{rotaru2024what}. 16 young, normal-hearing participants were instructed to listen to one out of two competing talkers located at $\pm 90^{\circ}$ w.r.t. the participant. This specific perceived spatial separation between speakers, presented through insert earphones, was obtained by convolving the two competing stimuli with the relevant head-related transfer functions. The stimuli consisted of science outreach podcasts in Dutch, for which videos were also available. The EEG was recorded using a $64$-channel BioSemi ActiveTwo system, while also $4$ electro-oculography (EOG) electrodes were used to record eye movements.
	
	The experiment consisted of $4$ different conditions with $2$ trials per condition, resulting in $8$ trials in total. Each trial was $\SI{10}{\minute}$ long. These $4$ different conditions, summarized in Table~\ref{tab:conditions}, implement different visual conditions such that the effect of eye gaze (movements) on AAD can be investigated. In each trial, the participant always had to listen to the to-be-attended speaker, who, after $\SI{5}{\minute}$, switched sides to emulate a spatial attention switch. Importantly, the randomization was done such that the $2$ trials from the same conditions were $\SI{40}{\minute}$ apart in time, resulting in a potentially substantial shift in statistics between both trials.
	
		\begin{table}[]
		\centering
		\begin{tabular}{@{}p{0.15\linewidth}p{0.19\linewidth}p{0.61\linewidth}@{}}
			\toprule
			\textbf{Condition} & \textbf{Auditory vs. \newline visual attention} & \textbf{Task}                                                                                    \\ \midrule
			No visuals         & Incongruent                            & Fixate on an imaginary point in the middle of the black screen while minimizing eye movements    \\ \hline
			Static video        & Congruent   & Fixate on a static video of the attended speaker located on the same side of the corresponding attended speech signal (left or right) \\ \hline
			Moving video       & Incongruent                            & Follow a moving video of the attended speaker along a random horizontal trajectory on the screen \\ \hline
			Moving target + noise & Incongruent & Follow a moving crosshair along a random horizontal trajectory on the screen. There is additional auditory background babble noise present at $\SI{-1}{\decibel}$ signal-to-noise ratio (SNR) \\ \bottomrule
		\end{tabular}
		\caption{An overview of the $4$ different conditions in the AV-GC-AAD dataset. Per condition, there are $2$ trials of $\SI{10}{\minute}$ long.}
		\label{tab:conditions}
	\end{table}
	
	A full description of the dataset can be found in Rotaru et al.~\cite{rotaru2024what}. This dataset is publicly available online~\cite{rotaru_2024_11058711}, minus the EEG data from three participants (subject 2, 5, and 6) who did not consent to making their data publicly available. For a few subjects, a particular trial or condition is not present (see \cite{rotaru_2024_11058711} for details), which is taken into account during evaluation. In this report, we use the version of the dataset that can be found online, so that all results are fully reproducible.

	\section{Methods}
	\label{sec:methods}
\begin{figure}
		\centering
		\includegraphics[width=\linewidth]{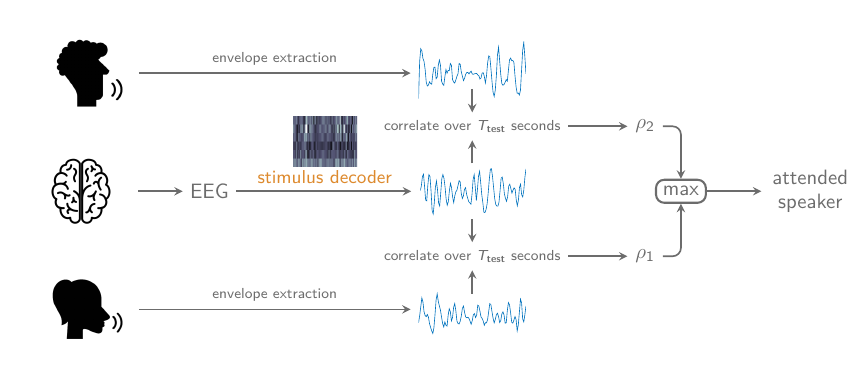}
		\caption{An overview of the linear stimulus decoding algorithm for AAD, in which the attended speech envelope is reconstructed from the neural responses and correlated with the presented speech envelopes to identify the attended one through the Pearson correlation coefficient. Based on Figure 3a in \cite{geirnaert2021eegBased} and Figure 2 in \cite{geirnaert2024fast}.}
		\label{fig:lsr}
	\end{figure}

	In this section, we revisit the basics of the linear stimulus decoding or reconstruction (backward modeling) approach for AAD. As shown in Figure~\ref{fig:lsr}, the stimulus decoding approach aims to reconstruct certain temporal features of the attended speech signal from the neural responses of the listener. A typical example of such a feature is the speech envelope, which we will also use in the remainder of this report. Other acoustic and linguistic features can also be used~\cite{diLiberto2015low,gillis2021neural,vanCanneyt2021enhanced}. This reconstruction can then be correlated with (features of) the presented competing speech signals to determine the attended speech signal~\cite{osullivan2014attentional,geirnaert2021eegBased}. 
	
	\subsection{Training the stimulus decoder}
	\label{sec:training}
	To decode or reconstruct the attended speech envelope $\hat{s}_a(t)$, with $t$ the time sample index, we linearly combine time-lagged copies of the EEG channels~\cite{geirnaert2021eegBased,osullivan2014attentional}:
	\begin{equation}
		\label{eq:spatio-temp-filt}
		\hat{s}_{a}(t) = \sum_{c = 1}^{C}\sum_{l = 0}^{L-1}d_c(l)x_c(t+l),
	\end{equation}
	where $x_c(t)$ is the value of the $c$-th EEG channel at time $t$, $d_c(l)$ is the $l$-th decoder coefficient for channel $c$, and $C$ and $L$ are the number of EEG channels and decoder time lags, respectively. As shown in~\eqref{eq:spatio-temp-filt}, the stimulus decoder $d_c(l)$ is an \emph{anti-causal} filter, where only time lags $l$ ranging from $0$ to $L-1$ \emph{after} the current stimulus sample at time $t$ are used, given we are trying to reconstruct the stimulus from the response. Using vector notations, \eqref{eq:spatio-temp-filt} can be rewritten as:
	\[
	\hat{s}_a(t) = \transpose{\vec{x}(t)}\vec{d},
	\]
	where $\vec{x}(t)$ contains all time lags per EEG channel:
	\[
	\vec{x}(t) = \begin{bmatrix}
		x_1(t) \\ \vdots \\ x_1(t+L-1) \\ x_2(t) \\ \vdots \\ x_C(t+L-1)
	\end{bmatrix} \in \R^{CL},
	\]
	and $\vec{d} \in \R^{CL}$ similarly stacks all spatio-temporal decoder coefficients $d_c(l)$. 
	
	To train the decoder coefficients $d_c(l)$, assume the availability of a training set of $T$ time samples, i.e., $\left\{\mat{X},\left(\vec{s}_{1},\vec{s}_{2}\right),\vec{y}\right\}$, with $\mat{X} = \begin{bmatrix} \mat{X}_{1} & \cdots & \mat{X}_{C}	\end{bmatrix} \in \R^{T \times CL}$, where $\mat{X}_{c} \in \R^{T \times L}$ is a Hankel matrix containing the time-lagged EEG data of the $c$\textsuperscript{th} channel:
	\[ \mat{X}_{c} = \begin{bmatrix}
		x_{c}(0) & x_{c}(1) & \cdots & x_{c}(L-1) \\
		x_{c}(1) & x_{c}(2) & \cdots & x_{c}(L) \\
		\vdots & \vdots &  & \vdots \\
		x_{c}(T-1) & 0 & \cdots & 0\\
	\end{bmatrix}.
	\]
	
	Furthermore, $\vec{s}_{1}$ and $\vec{s}_{2} \in \R^T$ contain the $T$ samples of both competing speech envelopes. Additionally, during training, we have knowledge of the attention labels $\vec{y} \in \R^T$, with $y(t) \in \{1,2\}$ indicating at each time sample which speech envelope corresponds to the attended envelope  $\vec{s}_{a} \in \R^T$, i.e.,
	\begin{equation}
		\label{eq:att-env}
		s_{a}(t)=\left\{
		\begin{array}{@{}*{2}{l}@{}}
			s_{1}(t) & \mbox{if $y(t) = 1$},\\
			s_{2}(t) & \mbox{if $y(t) = 2$}.
		\end{array}
		\right.
	\end{equation}
	
	The stimulus decoder can then be trained by minimizing the squared error between the reconstructed envelope $\hat{\vec{s}}_a = \mat{X}\vec{d}$ and attended one $\vec{s}_a$:
	\begin{equation}
		\label{eq:lsq}
		\hat{\vec{d}} = \underset{\vec{d}}{\text{argmin}} \norm{\vec{s}_{a}-\mat{X}\vec{d}}_2^2.
	\end{equation}
   The solution of~\eqref{eq:lsq} can be found by solving the normal equations, leading to:
	\[
	\hat{\vec{d}} = \inverse{\mat{R}}_{xx}\vec{r}_{xs},
	\]
	where
	\begin{equation}
		\label{eq:cov}
		\mat{R}_{xx} = \transpose{\mat{X}}\mat{X} \in \R^{CL \times CL}
	\end{equation}
	corresponds to the (unnormalized) EEG autocorrelation matrix and
	\begin{equation}
		\label{eq:rxs}
		\vec{r}_{xs} = \transpose{\mat{X}}\vec{s}_{a} \in \R^{CL}
	\end{equation}
	to the (unnormalized) cross-correlation vector between the EEG and attended speech envelope.
	
	\subsection{Using the stimulus decoder for AAD}
	\label{sec:testing}
	\noindent
	Given a new, unseen segment of EEG data $\mat{X}^{(\text{test})} \in \R^{T_{\text{test}}\times CL}$ and corresponding competing speech envelopes $\vec{s}_1^{\text{(test)}}$ and $\vec{s}_2^{\text{(test)}} \in \R^{T_{\text{test}}}$, the goal is to identify the attended speaker in this segment, i.e., to determine the correct label $y^{\text{(test)}}$. The trained decoder $\hat{\vec{d}}$ can now be used to reconstruct the speech envelope of the attended speaker $\hat{\vec{s}}_{a}^{\text{(test)}}=\mat{X}^{\text{(test)}}\hat{\vec{d}}$ from the EEG of the listener, which can then be compared with the competing speech envelopes through the Pearson correlation coefficient. The speaker that exhibits the highest correlation ($\rho(\hat{\vec{s}}_{a}^{\text{(test)}},\vec{s}_{1}^{\text{(test)}})$ or $\rho(\hat{\vec{s}}_{a}^{\text{(test)}},\vec{s}_{2}^{\text{(test)}})$) is identified as the attended speaker. This is summarized in Figure~\ref{fig:lsr}.
	
	A crucial parameter in determining the AAD accuracy (i.e., number of correct decisions) is the decision window length $T_{\text{test}}$, i.e., the number of samples used to compute the correlation. Given that the variability of these correlations increases with lower number of samples, the accuracy typically quickly decreases when using shorter decision window lengths~\cite{geirnaert2020interpretable}.
	
	\section{Experiments}
	\label{sec:experiments}
	In Section~\ref{sec:preprocessing}, we explain the preprocessing steps on the speech stimuli and EEG data. Section~\ref{sec:decoder} gives the details of the stimulus decoder settings, while Section~\ref{sec:evaluation} explains all procedures to evaluate the AAD stimulus decoding algorithm on the AV-GC-dataset. 
	
	\sloppy All MATLAB code to reproduce all experiments and results is available at \url{https://github.com/AlexanderBertrandLab/linear-stimulus-reconstruction-AAD-AV-GC-AAD-dataset}.
	
	\subsection{Preprocessing}
	\label{sec:preprocessing}
	To ensure full reproducibility, we start from the publicly online version of the dataset~\cite{rotaru_2024_11058711}. In the online version of the dataset, the EEG signals are already bandpass-filtered between $\SI{128}{\hertz}$ using a type II zero-phase Chebyshev filter and downsampled to $\SI{128}{\hertz}$. The speech envelopes are precomputed by applying a gammatone filterbank and computing the envelope per subband signal using a powerlaw operation with exponent $0.6$. All subband envelopes are then summed to one envelope, bandpass-filtered between $\SIrange{1}{40}{\hertz}$ and downsampled to $\SI{128}{\hertz}$ similarly to the EEG.
	
	We have, additionally, bandpass-filtered both EEG data and speech envelopes between $\SIrange{1}{9}{\hertz}$~\cite{das2016effect} using a $4$\textsuperscript{th}-order zero-phase Butterworth filter and downsampled them to $\SI{20}{\hertz}$. All EEG channels and speech envelopes were subsequently z-scored (mean and standard deviation put to $0$ and $1$, respectively) per $\SI{10}{\minute}$-trial.
	
	\subsection{Decoder settings}
	\label{sec:decoder}
	For the decoder, a filter range of $\SIrange{0}{400}{\milli\second}$ post-stimulus lags is chosen, such that $L = 9$ (at $\SI{20}{\hertz}$), while $C = 64$ given the $64$-channel EEG system.
	
	To avoid overfitting when only limited amounts of training data are available (e.g., in the subject-specific, per-condition evaluation (see below)), we use shrinkage to regularize the estimation of the EEG covariance matrix in \eqref{eq:cov}~\cite{geirnaert2021unsupervised,geirnaert2022timeAdaptive}. The regularization parameter is heuristically determined using the Ledoit-Wolf shrinkage estimator~\cite{ledoit2004well}, which is recommended as the state of the art~\cite{lotte2018review}.
	
	\subsection{Evaluation procedure}
	\label{sec:evaluation}
	Several deep learning approaches in AAD and neural tracking suffer from shortcuts and implicit trial-overfitting due to the uncareful validation procedure they employ~\cite{puffay2023relating}. Although the risk for (linear) stimulus reconstruction approaches that rely on correlation-based decisions is much lower, we will employ careful cross-validation to exclude all risks of overfitting to such potential shortcuts. We will use both subject-specific and subject-independent decoders, where the latter allows to assess the generalization abilities of the AAD algorithm in the hardest sense, i.e., across different subjects.
	
	In \emph{subject-specific} decoders, the stimulus decoder is trained and tested on data from the same subject. Such decoders generally perform better as they are tailored to the test subject. To avoid implicit overfitting, we will use a \emph{leave-one-trial-out} cross-validation (LOTO-CV) procedure, i.e., we always leave out a complete trial of $\SI{10}{\minute}$ from the training set. We will do this both within conditions and across conditions. Within-condition evaluation boils down to only taking the two trials corresponding to one of the four conditions in Table~\ref{tab:conditions} and performing LOTO-CV on those two trials. In this case, there is only $\SI{10}{\minute}$ of training data. Across-condition evaluation refers to LOTO-CV on all trials of all four conditions combined. Alternatively, we will also perform \emph{leave-one-condition-out} CV (LOCO-CV) across all conditions, i.e., we will leave out both trials from one condition in the training set. This allows to check generalization to specific conditions.
	
	In \emph{subject-independent} decoders, the stimulus decoders are trained on data from different subjects than the test subject, i.e., a \emph{leave-one-subject-out} CV (LOSO-CV) procedure is used. We will perform LOSO-CV across all conditions combined. Secondly, to refute arguments that there is insufficient training data to train more complex models, we will also train a stimulus decoder based on all subjects of the KU Leuven AAD dataset 2016~\cite{das2019dataset} (16 subjects $\times \SI{72}{\minute} \approx \SI{19}{\hour}$ of training data). This subject-independent decoder is then again tested on all subjects of the AV-GC-AAD dataset. The same preprocessing steps as in Section~\ref{sec:preprocessing} are applied on the KU Leuven AAD dataset 2016, which was recorded using the same EEG system.
	
	The significance level for the decoding accuracy is computed as the $95$-percentile of the inverse binomial cumulative distribution function with $50\%$ chance of success and the corresponding number of decisions.
	
	\section{Results and discussion}
	\label{sec:results-discussion}
	In Section~\ref{sec:ss-decoding}, we show and discuss the results of the experiments using a subject-specific decoder, while in Section~\ref{sec:si-decoding}, we show the results of subject-independent decoding.
	
	\subsection{Subject-specific decoding}
	\label{sec:ss-decoding}
	
\begin{figure}
		\centering
		\includegraphics[width=0.9\linewidth]{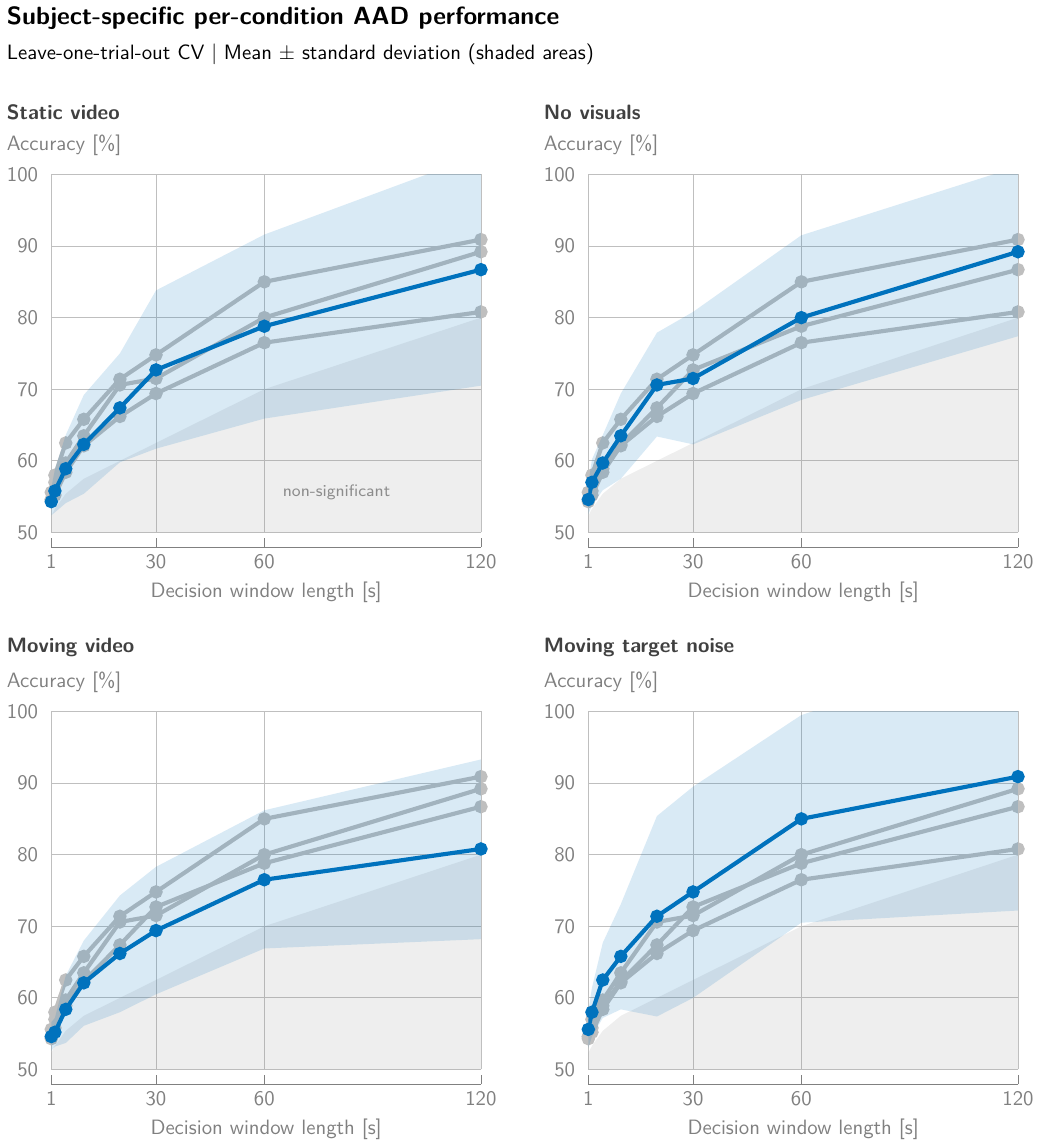}
		\caption{Using leave-one-trial-out CV for subject-specific decoding per individual condition leads to significant AAD performances for every single condition, even when the visual instruction is incongruent with the direction of auditory attention (moving video and moving target + noise). Gray lines are replicas per condition, provided as a reference.}
		\label{fig:ptau-SS-perCondition}
	\end{figure}

	Figure~\ref{fig:ptau-SS-perCondition} shows the results of the LOTO-CV on a per condition basis, while Figure~\ref{fig:SS-acrossConditions} shows the results across conditions. 
	
	\subsubsection{Per condition evaluation}
	
	For the per-condition LOTO-CV, one $\SI{10}{\minute}$-trial of each condition is used as training data, while the other one is used as the testing trial, and vice versa. The first crucial observation from Figure~\ref{fig:ptau-SS-perCondition} is that the stimulus decoding algorithm works for \emph{each} condition, as significant AAD accuracies are achieved for every single condition. This implies that the participants were generally able to focus their attention on the instructed speaker despite the various visual instructions. Even in the case of a visually moving target (as in the moving video or moving target + noise condition), significant performances are obtained. 
	
	When comparing the different conditions, the moving video condition is the worst one. This is no surprise, as it is one of the hardest conditions, given the continuously and randomly moving video, independent of the direction of the attended speaker. On the other hand, the moving target + noise also has a randomly moving target (albeit a crosshair) and results in the highest accuracies. In this moving target + noise condition, the difficulty of the moving crosshair could be offset by the added auditory background noise. Das et al.~\cite{das2018eegbased} observed an increasing performance with mild background noise at $\SI{-1.1}{\decibel}$ SNR and $180^{\circ}$ speaker separation w.r.t. no background noise, potentially due to the higher listening effort of the participant resulting in stronger neural tracking. The same effect could explain the higher accuracy in the moving target + \emph{noise} condition w.r.t. the other conditions without background noise, as it is recorded using the same settings as in Das et al.~\cite{das2018eegbased}. 
	
	However, per-condition accuracies are generally lower than expected in AAD (e.g., around $90\%$ on $\SI{60}{\second}$ decision windows in O'Sullivan et al.~\cite{osullivan2014attentional}). However, this lower accuracy can be attributed to the low amount of training data per condition ($\SI{10}{\minute}$), as the performance increases with LOTO-CV across all conditions (see later) and, for example, when using random $10$-fold CV ($= \SI{18}{\minute}$ of training data) based on shorter segments of $\SI{60}{\second}$ ($86.9\%$ on $\SI{60}{\second}$ decision windows for the no visuals condition).
	
	\subsubsection{Across conditions evaluation}
	
	\begin{figure}
		\centering
		\begin{subfigure}{\linewidth}
			\centering
			\includegraphics[width=0.625\linewidth]{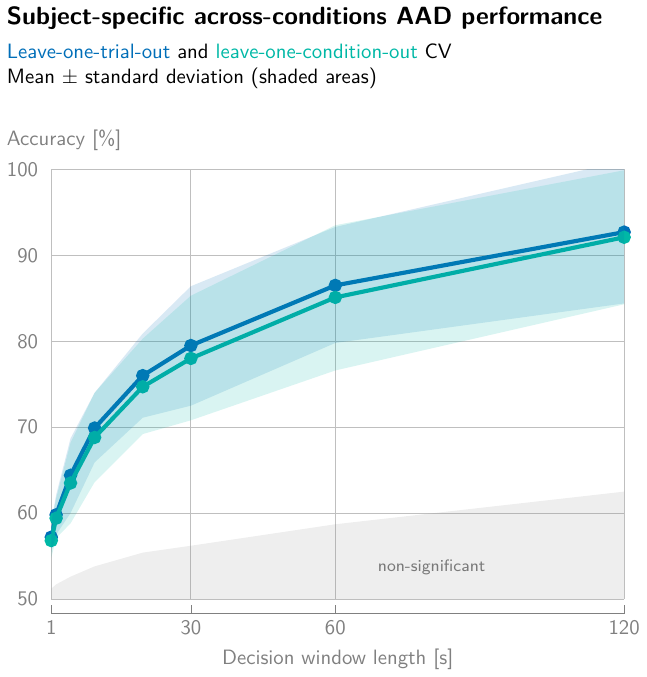}
			\caption{}
			\label{fig:ptau-SS-acrossConditions}
		\end{subfigure}
		
		\begin{subfigure}{\linewidth}
			\centering
			\includegraphics[width=0.9\linewidth]{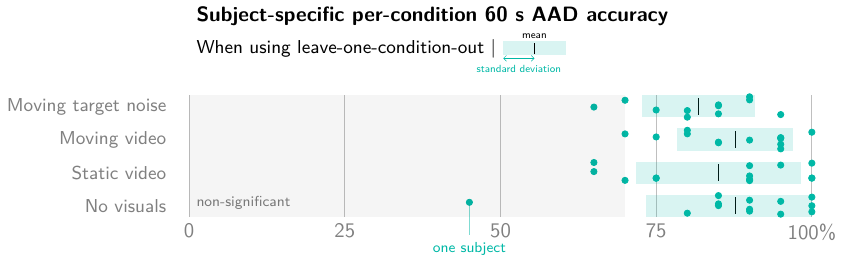}
			\caption{}
			\label{fig:swarm-SS-perCondition}
		\end{subfigure}%
		\caption{\textbf{(a)} Using leave-one-trial-out or leave-one-condition-out CV across all conditions at once both leads to significant and expected AAD performances. \textbf{(b)} A breakdown of the leave-one-condition-out CV accuracies per condition shows that generalization to every other condition is possible and similar. One dot represent the average $\SI{60}{\second}$ AAD accuracy for one subject.}
		\label{fig:SS-acrossConditions}
	\end{figure}
	
	Across conditions, both leave-one-trial-out and leave-one-condition-out CV are used. Figure~\ref{fig:ptau-SS-acrossConditions} shows that both CV evaluation schemes achieve very comparable and significant accuracies. This shows that generalization across all the different conditions using linear stimulus decoding is possible. Moreover, the accuracies are now comparable with other datasets (such as in O'Sullivan et al.~\cite{osullivan2014attentional} and Geirnaert et al.~\cite{geirnaert2021unsupervised}) as it achieves performances of $86.5\%$ (LOTO-CV) and $85.1\%$ (LOCO-CV) on $\SI{60}{\second}$ decision windows.
	
	To show that this generalization holds to every other condition, Figure~\ref{fig:swarm-SS-perCondition} shows a breakdown of the per-condition and per-subject AAD accuracy using $\SI{60}{\second}$ decision windows. All conditions have a similar mean accuracy around $85\%$ with a similar spread.
	
	\subsection{Subject-independent decoding}
	\label{sec:si-decoding}
	
	\begin{figure}
		\centering
		\includegraphics[width=.67\linewidth]{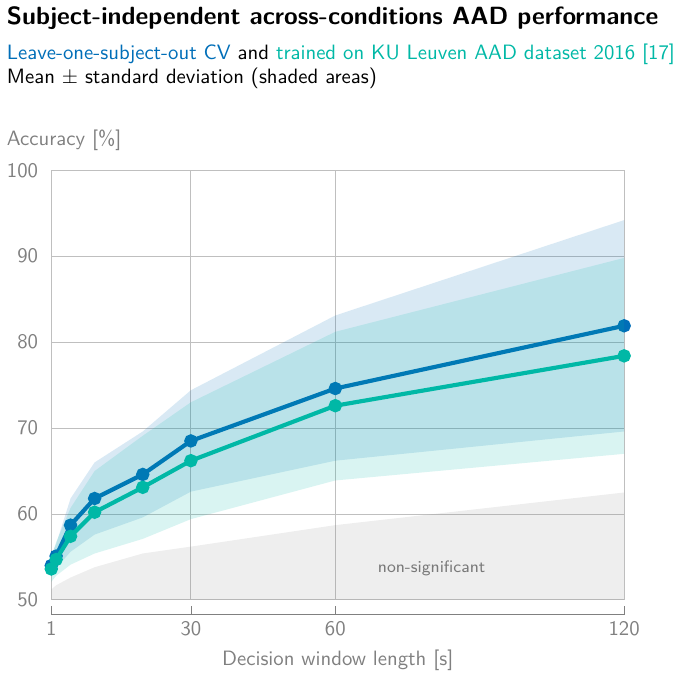}
		\caption{Using leave-one-subject-out CV and generalizing from the KU Leuven AAD dataset 2016 (i.e., subject-independent decoding) leads to significant AAD performances, showing that generalization across subjects and datasets is possible.}
		\label{fig:ptau-SI-acrossConditions}
	\end{figure}
	
	Figure~\ref{fig:ptau-SI-acrossConditions} shows the results of the leave-one-subject-out CV evaluation, where all data from one subject is entirely removed from the training set, and the generalization from the KU Leuven AAD dataset 2016 (i.e., subject-independent decoding). The significant AAD classification performance in both cases shows that even generalization across subjects and datasets with the linear stimulus decoder is possible. 
	
	For LOSO-CV, this good generalization is achieved despite the large variety of audiovisual conditions. A similar, significant performance is achieved when training the decoder on the KU Leuven AAD dataset 2016 on all conditions (on $\SI{60}{\second}$: $70.4\%$ on no visuals, $72.7\%$ on static video, $70.8\%$ on moving video, and $76.8\%$ on moving target + noise, respectively). This shows that even generalization across datasets to the AV-GC-AAD dataset is possible. The implication is that all the data of the KU Leuven AAD dataset 2016 could be additionally used to generate more training data for more complex models.
	
	\section{Conclusion}
	\label{sec:conclusion}
	In this short report, we have shown that linear stimulus reconstruction works on the KU Leuven audiovisual, gaze-controlled auditory attention decoding (AV-GC-AAD) dataset. Significant and expected performances are obtained for subject-specific decoding, both within all different audiovisual conditions and across all conditions. Generalizing across different subjects with a subject-independent decoder also yields significant AAD performance. Moreover, even generalizing from a different dataset to the AV-GC-AAD dataset is possible, opening up possibilities to generate more training data to train more complex models. The results show that the participants were able to follow the instructions and attend to the correct speaker and that their auditory attention is decodable. Moreover, we have provided a baseline analysis and performance for all future AAD algorithms on the AV-GC-AAD dataset, showing that above-chance accuracy is easily achievable on these data with simple linear models. Future algorithms should be able at least to beat the linear stimulus reconstruction algorithm on this dataset.

\section*{References}
\bibliographystyle{IEEEtran}
\bibliography{biblio}

\end{document}